\documentclass[a4paper,11pt]{article}
\usepackage{pos}

\title{Extensions to $\Lambda$CDM at Intermediate Redshifts to Solve the Tensions ?}
\ShortTitle{Extensions to $\Lambda$CDM at Intermediate Redshifts}

\author[\,a,b,c]{Ziad Sakr}

\affiliation[a]{Institut fur Theoretische Physik, Philosophenweg 16, Heidelberg, Germany}
\affiliation[b]{IRAP, Université de Toulouse, CNRS, CNES, UPS, Toulouse, France}
\affiliation[c]{Faculty of Sciences, Universit\'e St Joseph; Beirut, Lebanon}

\emailAdd{sakr@thphys.uni-heidelberg.de}\emailAdd{ziad.sakr@net.usj.edu.lb}

\abstract{Models of dark energy or modified gravity that tries to alleviate the tensions on the Hubble constant ($H_0$) and the matter fluctuation parameter ($\sigma_8$) are usually parameterized as function of either late or early time cosmic evolution. In this work we rather focus on one that could privilege extensions to $\Lambda$CDM on intermediate redshifts by mean of a Gaussian-like window function with a free moving centre $a_{Gwin}$ combined with a modified gravity parameter $\mu_{Gwin}$ and an extension of the equation of state parameter $\omega_{Gwin}$. Using different combinations of the latest available current datasets subject of the discrepancies, such as the cosmic microwave (CMB) background power spectrum, the baryonic acoustic scale (BAO) in galaxy distribution, Weak lensing (WL) shear and galaxy clustering cross correlations and local hubble constant measurements, we investigate whether such model could alleviate each or both $H_0$ and $\sigma_8$ tensions. We found when combining all probes that the $\sigma_8$ tension is alleviated while the $H_0$ is reduced with a small preference for a positive $\omega_{Gwin}$ without a particular preference for a redshift or a $\mu_{Gwin}$ different from its equivalent $\Lambda$CDM value. However, if we follow another approach and compare the two sets of the probes subject of discrepancy i.e. CMB+BAO vs WL+local $H_0$, we found that the model is able of solving the $\sigma_8$ discrepancy at the expense of a enlargement of the constraints, while the Hubble constant discrepancy is not that affected due to the fact that the two likelihood contours are stretched in parallel directions. We conclude that modifying $\Lambda$CDM cosmology at intermediate redshifts within our model, and the constraints from the datasets used in this study, are not likely a viable solution to solve both tensions.}

\FullConference{%
  Corfu Summer Institute 2022 "School and Workshops on Elementary Particle Physics and Gravity",\\
  28 August - 1 October, 2022\\
  Corfu, Greece}

\begin{document}
\maketitle

\section{Introduction}\label{sec:sample1}

The precision with which the cosmological parameters were measured by precise observations such as the cosmic microwave background temperature and polarisation power spectrum \cite{Planck:2018vyg}, the baryonic acoustic oscillation signature feature in galaxy distributions \cite{eBOSS:2020yzd} or the weak lensing cosmic shear correlations \cite{DES:2021wwk,KiDS:2020ghu} has drastically improved since the end of the last century reaching the percent level. However, this has also resulted in the appearance of some tensions on their values when measured by different probes, notably the Hubble constant $H_0$ when inferred from CMB +BAO with respect to its value determined from local distance of Cepheids stars \cite{2022ApJ...934L...7R}, and to a lesser degree the matter fluctuation parameter $\sigma_8$ \cite{Sakr:2018new} when its value derived from CMB is compared to that obtained from near redshift weak lensing measurements or galaxy cluster abundances \cite{KiDS:2020ghu,2016A&A...594A..24P}.
Apart from possible misdetermination of the systematics involved, many theoretical solutions were proposed to solve these tensions, starting first with models based on a late time modification of $\Lambda$CDM growth or the dark energy equation of state parameter $\omega$. Most of these attempts to solve, either $H_0$ \citep{Zhao:2017cud,Li:2019yem,Raveri:2019mxg,Braglia:2020iik,DiValentino:2019jae,Banihashemi:2020wtb} or $\sigma_8$ \cite{Joudaki:2017zdt,Sakr:2021jya} failed, in particular when data from the baryonic acoustic oscillations (BAO) were included \cite{Jedamzik:2020zmd,Sakr:2021jya}, because the later strongly tie the sound horizon signature in  CMB oscillations at early redshift to those imprinted in late times galaxy distributions. But also early modifications that try to reduce the sound horizon at the last scattering surface without messing with its value at the BAO level \citep{Poulin:2018cxd,Knox:2019rjx} turns out to exasperate the $\sigma_8$ tension as shown for example in \cite{Hill:2020osr} or \citep{Schoneberg:2021qvd}.There have also been many other proposed solutions to solve one or both tensions, such as modifications of the gravitational coupling to matter e.g. \cite{Khosravi:2017hfi,Sakr:2021nja} , dark matter evolution \citep{Naidoo:2022rda}, primordial magnetic fields \citep{Jedamzik:2020krr} or interacting dark matter dark energy models \cite{Yang:2018uae,Gomez-Valent:2020mqn}.  All failed to substantially alleviate both tensions at the same time. This has lead us in this work to explore modifications to $\Lambda$CDM model at the modified gravity model or dark energy equation of state parameter level, that rather privilege deviation at the intermediate redshifts, i.e. around but not restricted to $z\sim1$ by mean of a Gaussian-like filter which parameters are, its width and the redshift value of its centre. This is also motivated by the fact that current datasets have not yet fully explored this range of redshifts, between $z\sim1$ and $2$, with a large coverage of the sky and high density of the detected point sources at the same time, similar to what was already done at low redshifts e.g. \cite{eBOSS:2020yzd,DES:2021wwk} but also to a lesser extent around or a little above z $\sim2$ \cite{duMasdesBourboux:2017mrl,Chabanier:2018rga}. Thus, there might be still room for our model's parameters to explore values that could alleviate the tensions. In order to asses the ability of our approach in reducing or not the discrepancies, we shall use different datasets combinations of current observations, focusing on the ones subject to the aforementioned tensions and confront them to the theoretical predictions of our model. This report is organised as follows: In Sect.~\ref{sect:modelanddata} we present the equations used to describe a model favouring deviation at intermediate redshifts to $\Lambda$CDM, and review the datasets and pipeline used for the model's parameter estimation. In Sect.~\ref{sect:results}, we show and discuss the results before drawing our conclusions in Sect.~\ref{sect:conclusion}

\section{Model and datasets}
\label{sect:modelanddata}
The evolution of perturbations in modified gravity could be described by the following relations between the time and scale potentials and the two modified gravity parameterization $\mu(a,k)$ and $\eta(a,k)$ :

\begin{align}
 -k^2\,\Psi(a,\vec{k}) &= \frac{4\pi\,G}{c^4}\,a^2\,\bar\rho(a)\,\Delta(a,\vec{k})\times\mu(a,k), \label{eq:mu}\\ 
 \Phi(a,\vec{k}) &= \Psi(a,\vec{k})\times\eta(a,k),    \label{eq:eta}\\
 \end{align}
where $\bar\rho\Delta=\bar\rho\delta+3(aH/k)(\bar\rho+\bar p)v$ is the comoving density perturbation of $\delta=(\rho-\bar{\rho})/\bar{\rho}$, and $\rho$, $p$ and $v$ are, respectively, the density, pressure and velocity with the bar denoting mean quantities. 
$\Phi$ and $\Psi$ are the Bardeen potentials entering the perturbed FLRW metric, which in Newtonian gauge is
\begin{equation}
ds^2=a^2\left[-(1+2\Psi)d\tau^2+(1-2\Phi)d\vec{x}^2\right]\,.
\end{equation}
The two functions $\mu(a,k)$, and $\eta(a,k)$ encode the possible deviations from GR. Here we consider a parameterisations where each variation is restricted by mean of a Gaussian-like window (similar selection was also suggested in \cite{Denissenya:2017uuc}) which parameters are function of the scale factor $a$ following:

\begin{align}
 \mu &= 1 + \mu_{Gwin} \, e^{-{\left(\frac{a-a_{Gwin}}{\Delta a_{Gwin}}\right)}^2}, \label{eq:mugauss}\\ 
 \eta &= 1 + \eta_{Gwin} \, e^{-{\left(\frac{a-a_{Gwin}}{\Delta a_{Gwin}}\right)}^2}, \label{eq:sigmagauss}
\end{align}
We also introduce a variation in the dark energy equation of state formulation following:

\begin{equation}
\omega = \omega_0 + \omega_{Gwin} \, e^{-{\left(\frac{a-a_{Gwin}}{\Delta a_{Gwin}}\right)}^2}
\end{equation}
%
\begin{figure}
	\includegraphics[width=0.5\columnwidth]{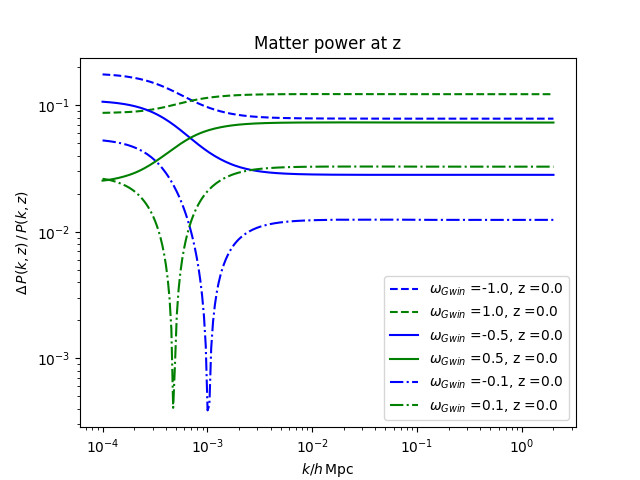}
    \includegraphics[width=0.5\columnwidth]{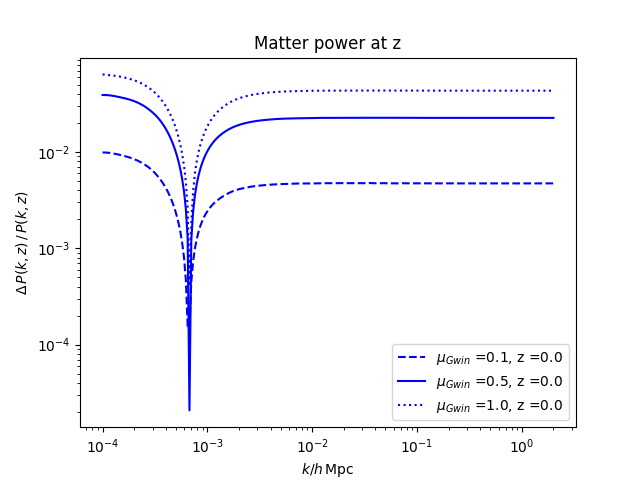}
    \includegraphics[width=0.5\columnwidth]{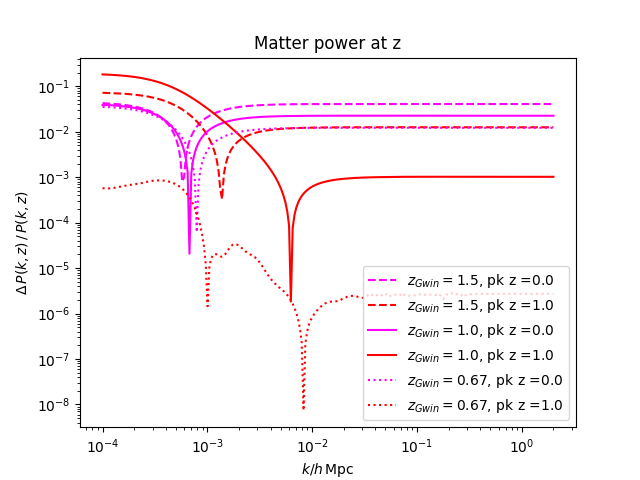}
    \includegraphics[width=0.5\columnwidth]{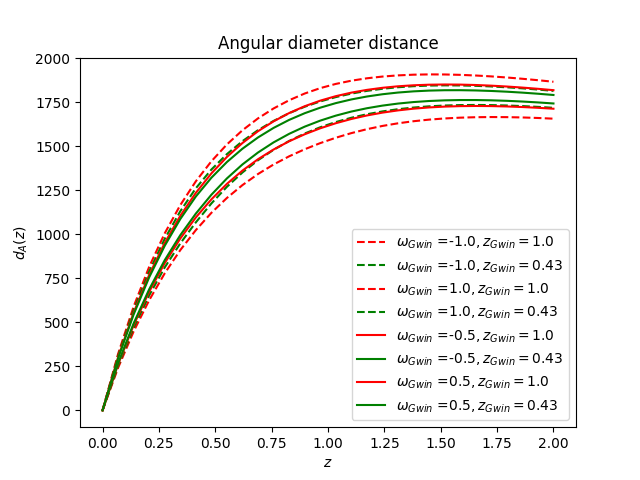}
    \caption{Showing the impact from varying the parameters of the $\rm{MG}_{\,z,win}$ model on the matter linear power spectrum with respect to that obtained within $\Lambda$CDM starting from the $\omega_{G win}$ parameter (left upper panel), then $\mu_{G win}$ (right upper panel), to end with $z_{G win}$ (equivalent to $a_{G win}$) parameter (bottom left panel). Showing as well the impact on the angular diameter distance from a variation of $\omega_{G win}$ and $z_{G win}$ (bottom right panel). When not varied, $\omega_{G win}$ is fixed to zero while $\mu_{G win}$ to 0.5 and $z_{G win}$ to 1.}
    \label{fig:GwinImpact}
\end{figure} 
To illustrate some of the phenomenological effects of the $\rm{MG}_{\,z,win}$ model's parameters on the angular diameter distance or the matter power spectrum ($P_k$), with the two being basic ingredients for many observables, we show in Fig.~\ref{fig:GwinImpact} the residual of $P_k$ with respect to $\Lambda$CDM model when varying our parameters, 
with the idea to check the effects when the window is centred around a redshift, on the aforementioned quantities at other redshifts. Starting with $\omega_{G win}$ in the left upper panel, we see that the difference increase along with the value of $\omega_{G win}$ to reach 10\% for $\omega_{G win}\sim1$, but also we observe that the result differs whether $\omega_{G win}$ is positive or negative and the behaviour switches between small and high scales, while for $\mu_{Gwin}$ in the right upper panel, the impact is more straightforward with a direct increase for $P_k$ on all scales with the value of $\mu_{Gwin}$. Finally we show in the bottom the effect from varying the centre of the window parameter, $a_{Gwin}$, a distinct feature of our model, on the power spectrum and the angular diameter distance when also varying $\omega_{G win}$ in the geometrical quantity case. We see first for $P_k$ a rich non monotonic phenomenology with however a general behaviour showing that the impact of the window filter is affecting basically the observable the farthest its redshift is from the model's chosen window as seen for the magenta coloured lines, while for the effect on the background, the impact from varying $a_{Gwin}$ is to enhance or limit that induced from $\omega_{G win}$ with the most happening when cutting at higher redshift (red in comparison to the green lines). Thus, from the three parameter we showed, only $\mu_{Gwin}$ effect is overall monotonic but it is still necessary because it changes the amplitude of $P_k$ and, when combined with a variation of $a_{Gwin}$, it will as well further enhance the effect of the latter parameter. With already such richness in the phenomenology, we shall next perform a Bayesian study varying only these three parameters and leave $\eta_{Gwin}$ to be addressed by future studies.
For that, we use CMB temperature, polarization, their cross correlations C$_\ell$ and lensing spectrum D$_\ell$ likelihood and data of from Planck 2018 (Plk18) releases \cite{Planck:2019nip,Planck:2018vyg} to run Monte Carlo Markov Chains (MCMC) shown in the next section. We also include background observations from BAO measurements \cite{Beutler:2011hx,Ross:2014qpa,BOSS:2016wmc}  and combine with 3$\times$2  galaxy lensing, clustering and their cross correlated spectrum from DES collaboration \cite{DES:2017myr,DES:2021wwk} where we limit and cut to the linear scales.
We run our MCMC using \texttt{MGCLASS II} \cite{Sakr:2021ylx} which was further modified to include our model and interfaced with the cosmological data analysis codes e.g. \texttt{MontePython} \cite{2013JCAP...02..001A} in which we implemented the DES likelihood based on the collaboration's one. 

\section{Results}
\label{sect:results}

\begin{figure}
	\includegraphics[width=\columnwidth]{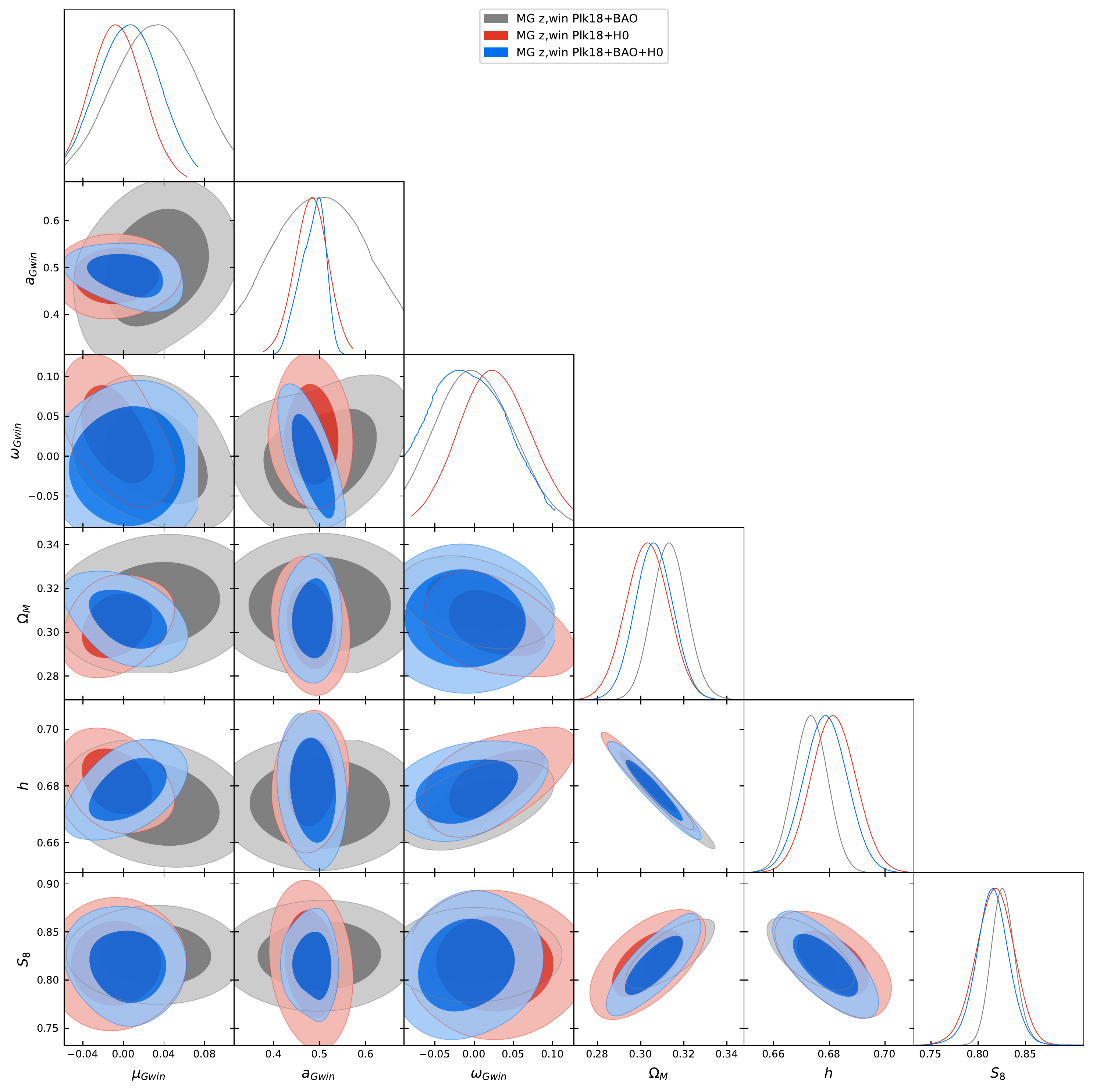}
    \caption{68\% and 95\% confidence contours for the parameters $\mu_{G win}$, $a_{G win}$, $\omega_{G win}$, $\Omega_m$, $h$ and $S_8$, inferred from different combinations of CMB $C_{\ell}^{TT,TE,EE}$+lens from Plk18, BAO measurments and local $H_0$ prior within the $\rm{MG}_{\,z,win}$ model allowing extensions to $\Lambda$CDM at intermediate redshifts.}
    \label{fig:CmbBaoHo}
\end{figure}

In Fig.~\ref{fig:CmbBaoHo} we show the MCMC inferred values for our model, while fixing $\eta_{Gwin}$ to zero and the width of the window $\Delta a_{G win}$ to 0.1, with a flat prior on $a_{G win}\in [0.3,0.7]$, starting from a combination of CMB and the BAO distance probe, since the two are in general in agreement on the values of the cosmological parameters. We observe (gray lines) that the model parameters are allowed to vary without however a substantial impact on the shift or the widening of the constraints on $H_0$ or $S_8={(\Omega_m/0.3)}^{0.5}\sigma_8$ parameters \footnote{We show here $S_8$ rather than $\sigma_8$ because it was adopted by the DES collaboration but also is what is effectively measured by weak lensing correlations.}. If we choose to combine CMB with $H_0$ priors from SH0ES, we observe that $S_8$ and $H_0$ are slightly shifted towards values compatible with those from discrepant probes with CMB, thus helping in reducing the tensions, and that the $\omega_{G win}$ is preferring a positive value, an intermediate redshift around $z\sim 1$ and negative values for $\mu_{G win}$. However, adding again BAO to the previous combination reverts back $\omega_{G win}$ and $\mu_{G win}$ to their equivalent $\Lambda$CDM values although keeping preference to $z\sim1$ as well as allowing a slight reduction of tensions on $H_0$ and $S_8$.\\

\begin{figure}
	\includegraphics[width=\columnwidth]{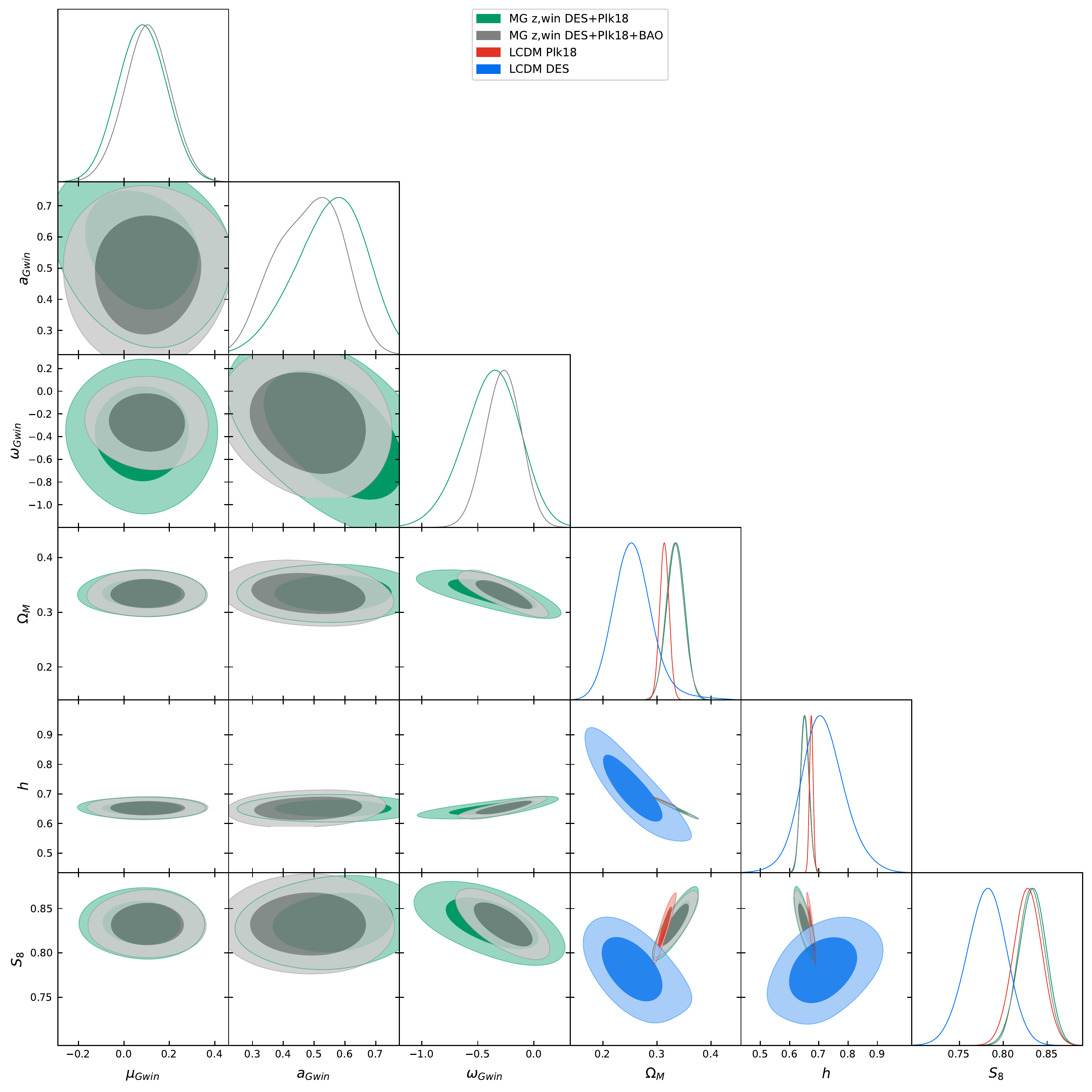}
    \caption{68\% and 95\% confidence contours for the parameters $\Omega_m$, $h$ and $S_8$, inferred from $C_{\ell}^{TT,TE,EE}$+lens from Plk18 or DES 3$\times$2pt within $\Lambda$CDM, in addition to $\mu_{G win}$, $a_{G win}$ and $\omega_{G win}$ within the $\rm{MG}_{\,z,win}$ model allowing extensions to $\Lambda$CDM at intermediate redshifts, in combination or not with BAO measurements in the latter case.}
    \label{fig:CmbBaoDesHo}
\end{figure}

We perform next the same MCMC but using the 3$\times$2pt probe from DES instead of that from local $H_0$. We observe in Fig.~\ref{fig:CmbBaoDesHo}, where we also show constraints within $\Lambda$CDM model for DES and CMB from Plk18 separately for comparison, that combining DES with CMB with or without BAO does not show preference for any redshift window centre value but prefers however negative values for $\omega_{G win}$ and positive ones for $\mu_{G win}$, thus opposite behaviour with respect to the case with local $H_0$ priors. But that has not the desired effect on the cosmological parameters subject of discrepancy since $H_0$ is shifted towards even smaller values below $\sim$ 0.675 while $S_8$ is clearly still within the CMB ones.\\

\begin{figure}
	\includegraphics[width=0.65\columnwidth]{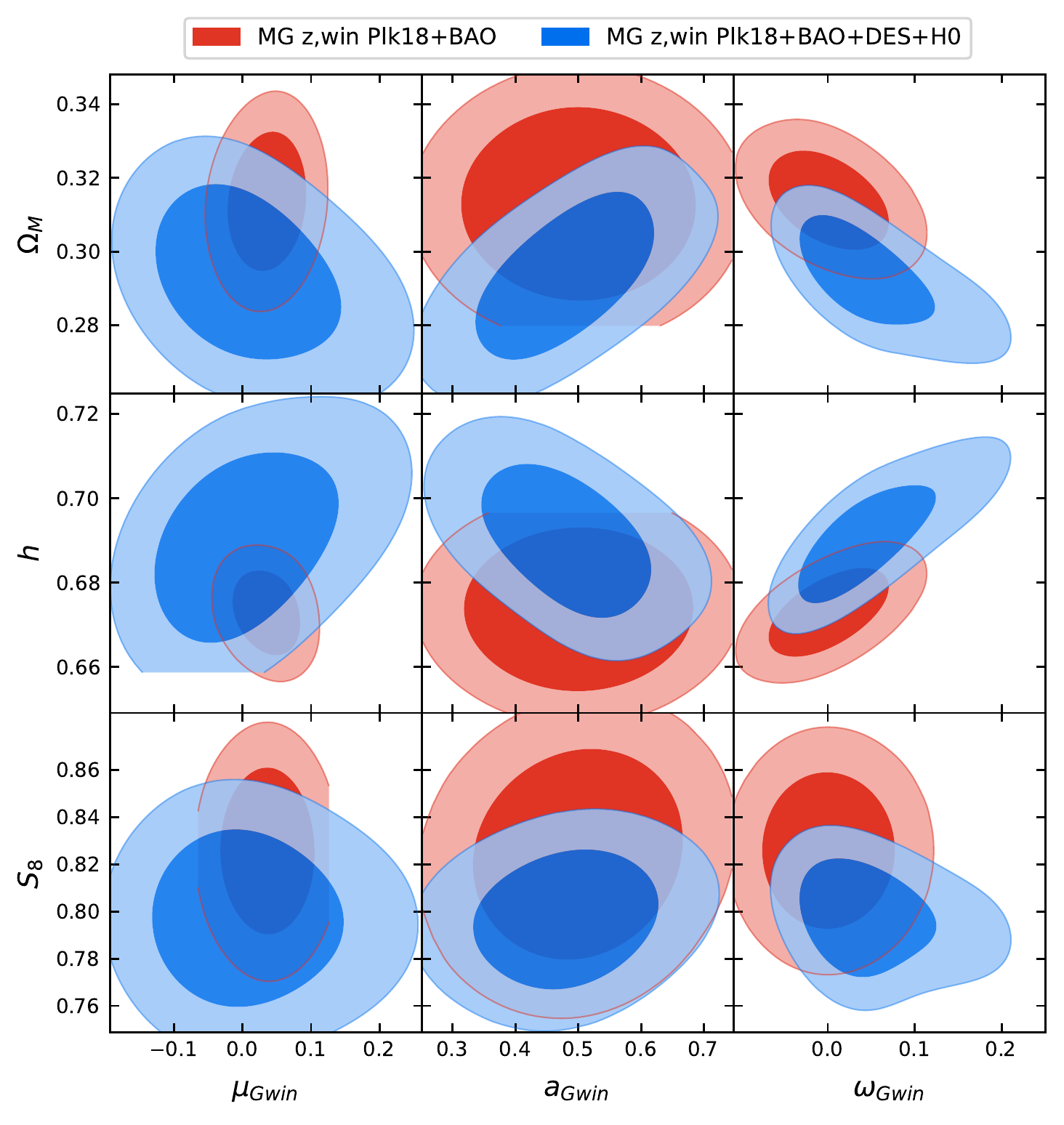}
    \caption{68\% and 95\% confidence contours for the parameters $\mu_{G win}$, $a_{G win}$, $\omega_{G win}$, $\Omega_m$, $h$ and $S_8$, inferred from combinations of CMB $C_{\ell}^{TT,TE,EE}$ + lens from Plk18 and BAO measurements in comparison to the same inferred from further combining with local $H_0$ prior and 3$\times$2pt and galaxy correlations and cross correlations from DES survey within the $\rm{MG}_{\,z,win}$ model allowing extensions to $\Lambda$CDM at intermediate redshifts.}
    \label{fig:zgwinplkBAOvsDESH0_1}
\end{figure}

\begin{figure}
	\includegraphics[width=\columnwidth]{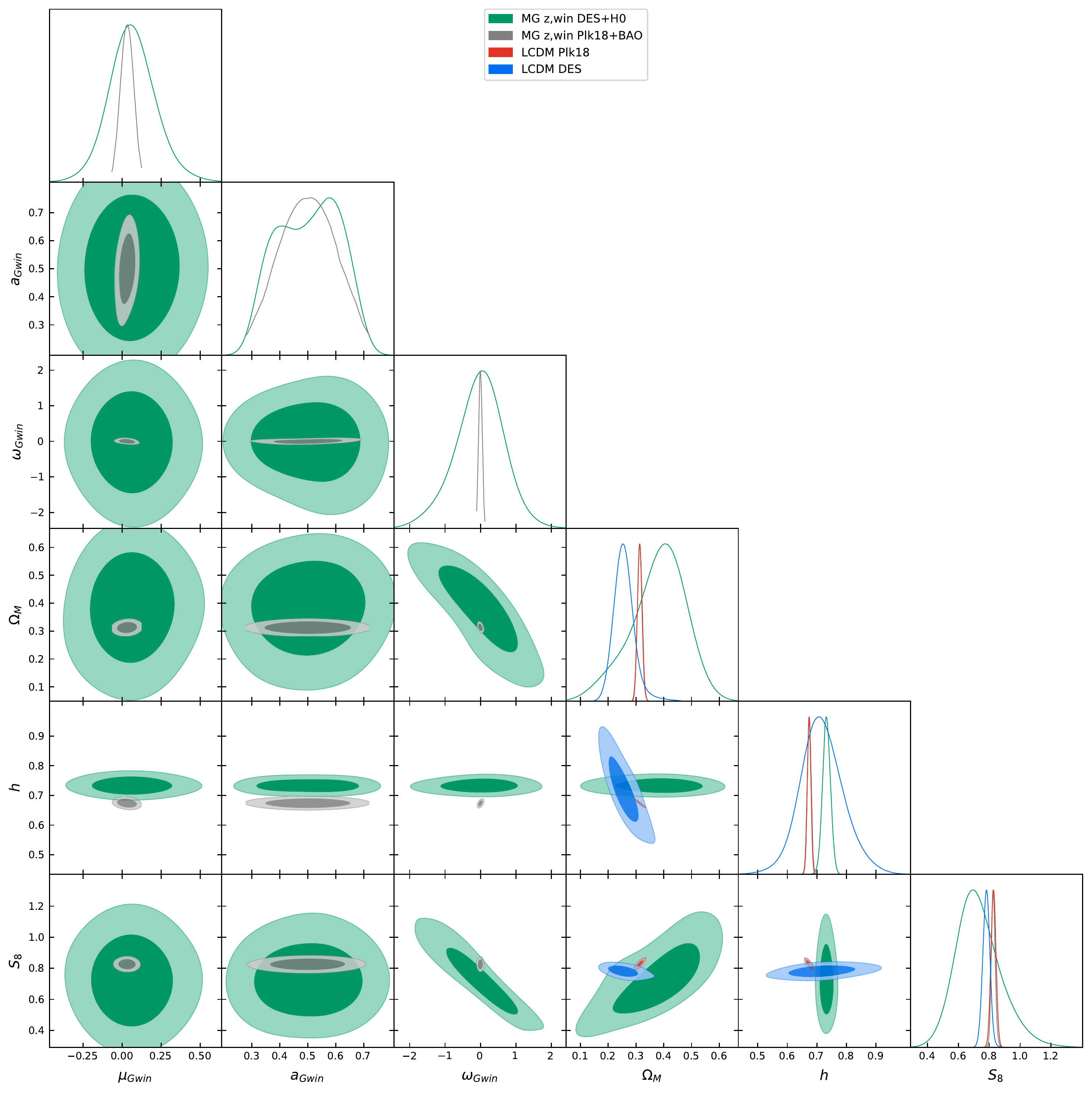}
    \caption{68\% and 95\% confidence contours for the parameters $\mu_{G win}$, $a_{G win}$, $\omega_{G win}$, $\Omega_m$, $h$ and $S_8$, inferred from combinations of CMB $C_{\ell}^{TT,TE,EE}$ +lens from Plk18 and BAO measurements in comparison to the same inferred from a separate combination of local $H_0$ prior and 3$\times$2pt and galaxy correlations and cross correlations from DES survey, all within the $\rm{MG}_{\,z,win}$ model allowing extensions to $\Lambda$CDM at intermediate redshifts.}
    \label{fig:zgwinplkBAOvsDESH0_2}
\end{figure}

The previous results make us expect what we would obtain from a combination of all the above probes. Nevertheless, and because the interplay from all the model's parameters could yield sometimes different constraints from those obtained from each data or subset of data alone, we show in Fig.~\ref{fig:zgwinplkBAOvsDESH0_1} MCMC posteriors from the combination of all probes in comparison to the CMB + BAO baseline. We observe first, as already noted, that the baseline does not show preference for our model despite a small positive shift for $\mu_{G win}$, and a trend previously found in the constraints from local lensing and clustering data with respect to those from $H_0$ local datasets ones, showing no preference for our model parameters for values different from their equivalent $\Lambda$CDM ones except a positive preference for $\omega_{G win}$. However, we observe that the bounds on $H_0$ and $S_8$ are shifted towards values compatible with those obtained from DES and SH0ES alone with stronger reduction for $S_8$. This leads us to perform a final MCMC test from separate combinations of the two discrepant probes each to try to better understand the reason behind such behaviour. In this case, we observe in Fig.~\ref{fig:zgwinplkBAOvsDESH0_2} that our proposed extension of $\Lambda$CDM is not able of alleviating the $H_0$ tension due to the fact that the two 2D contours resulting from CMB+BAO (gray lines) remain distant from those obtained from DES+SH0ES combination (green lines) since they both extend in directions parallel to each others, while the tension on $S_8$ is alleviated at the cost of a large widening of the constraints, which has less statistical evidence than e.g. a shift in the discrepant parameters solving the discrepancy without a substantial loss in the constraining power. We deduce that the reduction of tension seen in the previous case when combining all probes is not reliable since their parameters are still showing discrepancy when considered separately.

\section{Conclusion}
\label{sect:conclusion}

In this work we considered a model that extends $\Lambda$CDM at intermediate redshifts by mean of three parameters $\mu_{G win}$,  $\omega_{G win}$ and $a_{G win}$, with the first encapsulating modified gravity theories that change the growth of structures while the second affects the equation of state dark energy parameter while both entering through a multiplication by a Gaussian window centred at redshift value equal to $a_{G win}$. The aim was to test whether the $H_0$ and $\sigma_8$ (or the adopted $S_8$ in this work) tensions can be alleviated by privileging redshifts at epochs near $z\sim1$ where the current data is still not stringiest enough to constrain such extensions, and that by performing a Bayesian study using different combinations of CMB, BAO, local $H_0$ priors and 3$\times$2pt clustering and lensing galaxies probes. 
Combining CMB and local $H_0$ priors, we found that $z\sim1$ is preferred with a positive value for $a_{G win}$ and a negative one for $\mu_{G win}$ while the opposite is observed when combining instead with 3$\times$2pt probe, with a small reduction of the two tensions for the first case. When combining with BAO in each cases we found that the intermediate epoch is still privileged but $\omega_{G win}$ and $\mu_{G win}$ revert to their null $\Lambda$CDM values. Finally, combining all the probes only show preference for a positive value for $\omega_{G win}$ with nevertheless a reduction to the $H_0$ and $S_8$ tensions. However, when we compared separate combination of CMB+BAO versus one with $H_0$+3$\times$2pt probes, we found that our model has not the ability to fix the Hubble tension since the two discrepant contours extend in parallel directions while it could solve the $S_8$ tension at the price of enlarging the inferred constraints. We conclude that a modification of $\Lambda$CDM at intermediate redshifts is not able of solving the discrepancy on $H_0$ and $\sigma_8$ parameters.

\acknowledgments

Z.S. acknowledges funding from DFG project 456622116 and support from the IRAP Toulouse and IN2P3 Lyon computing centres.

\bibliographystyle{JHEP}
\bibliography{interm_cosmo}



\end{document}